\documentclass[noshowpacs,amsmath,twocolumn,superscriptaddress,8pt,aps,prb]{revtex4-1} % this article uses the Revtex 4-1 format
\usepackage{setspace}
\usepackage{amsmath}
\usepackage{braket}
\usepackage{breqn}
\usepackage{graphicx}
\usepackage[nearskip,margin = 0pt]{subfig}
\usepackage{verbatim} % for commenting
\usepackage{amsfonts}
\usepackage{amssymb}
\usepackage{epstopdf}
\usepackage{physics}
\usepackage{xcolor}
\usepackage{ragged2e}
\usepackage{array}
\usepackage[resetlabels,labeled]{multibib}

\DeclareGraphicsExtensions{.pdf,.eps,.png,.jpg,.mps}

\begin{document}

\title{Quantum Optics of Soliton Microcombs}
\author{Melissa A. Guidry$^{*,1}$, Daniil M. Lukin$^{*,1}$, Ki Youl Yang$^{*,1}$, Rahul Trivedi$^{1,2}$ and Jelena Vu\v{c}kovi\'{c}$^{\dagger,1}$\\
\vspace{+0.05 in}
$^1$E. L. Ginzton Laboratory, Stanford University, Stanford, CA 94305, USA.\\
$^2$Max-Planck-Institute for Quantum Optics, Hans-Kopfermann-Str. 1, 85748 Garching, Germany
}

\begin{abstract}

Soliton microcombs --- phase-locked microcavity frequency combs --- have become the foundation of several classical technologies in integrated photonics, including spectroscopy, LiDAR, and optical computing. Despite the predicted multimode entanglement across the comb, experimental study of the quantum optics of the soliton microcomb has been elusive. In this work, we use second-order photon correlations to study the underlying quantum processes of soliton microcombs in an integrated silicon carbide microresonator. We show that a stable temporal lattice of solitons can isolate a multimode below-threshold Gaussian state from any admixture of coherent light, and predict that all-to-all entanglement can be realized for the state. Our work opens a pathway toward a soliton-based multimode quantum resource. 

\end{abstract}

\maketitle

Kerr optical frequency combs \cite{kippenberg:2018:Review, diddams:2020:Review, Kues:2019:NaturePhotonics} are multimode states of light generated via a third-order optical nonlinearity in an optical resonator. When a Kerr resonator is pumped weakly, spontaneous parametric processes populate resonator modes in pairs. In this regime, Kerr combs can be a quantum resource for the generation of heralded single photons and energy-time entangled pairs\cite{grassani2015micrometer, Weiner:2017:Optica, Moody:2020:AlGaAs}, multiphoton entangled states\cite{Morandotti:2016:Science, kues2017chip, Samara:2020:Swapping}, and squeezed vacuum \cite{vaidya2020broadband, zhao2020near, arrazola2021quantum}. When pumped more strongly, the parametric gain can exceed the resonator loss and give rise to optical parametric oscillation (OPO) and bright comb formation. The modes of a Kerr comb can become phase-locked to form a stable, low-noise dissipative Kerr soliton (DKS). This regime of Kerr comb operation has become the foundation of multiple technologies, including comb-based spectroscopy\cite{Suh:2016:Science}, LiDAR\cite{Riemensberger:2020:Nature}, optical frequency synthesizers\cite{DODOS:2018:Nature}, and optical processors\cite{Feldmann:2021:Nature}. 

While the soliton microcomb has been mostly studied classically, it is nonetheless fundamentally governed by the dynamics of quantized parametric processes: each resonator mode is coupled to every other mode through a four-photon interaction. The multimode coupled processes of the DKS resemble those of the well-studied synchronously pumped OPO\cite{roslund2014wavelength, cai2017multimode}. 
If the quantum processes can be harnessed, soliton microcombs may open a pathway toward the experimental realization of a multimode quantum resource\cite{cai2017multimode, Pfister:2014:60modes, pfister2019continuous} in a scalable, chip-integrated platform\cite{wang2020integrated, wu2020quantum}.
However, the quantum-optical properties of soliton microcombs have not yet been directly observed: the first glimpse into non-classicality was the recent measurement of the quantum-limited timing jitter of the DKS state\cite{matsko2013timing, bao2021quantum}. 

Linearization\cite{chembo2016quantum} --- the formal separation of the optical state into the mean field solution and the quantum fluctuations --- is one approach to simplify the otherwise intractable many-body quartic Hamiltonian of the DKS. 
The result of this approximation is a quadratic Hamiltonian, where the classical amplitudes $A_m$, obtained for example through the spatiotemporal Lugiato-Lefever equation (LLE)\cite{chembo::2013::LLE}, drive the spontaneous parametric processes between quantum modes $\hat{a}_j$:

\begin{equation}
\hat{H}_{\text{int}} =  -\frac{g_0}{2} \sum_{m,n,j,k} \delta_{\text{FWM}} ( A_m A_n \hat{a}_{j}^\dag \hat{a}_{k}^\dag + A_k^* A_n \hat{a}_{j}^\dag \hat{a}_{m} + \text{h.c.} )
\label{eq:H}
\end{equation}
where $\delta_{\text{FWM}} = \delta_{(j+k-m-n)}$ is the four-wave mixing (FWM) mode number matching condition and $g_0$ is the nonlinear coupling coefficient. 
This approximation has successfully modelled the quantum correlations developed by a coherent soliton pulse after propagating through a fiber \cite{haus1990quantum, spalter1998observation}, where the Kerr interaction is weak. The DKS state, however, is itself generated by strong Kerr interactions where linearization can break down and yield unphysical predictions\cite{navarrete2014regularized, vernon2015strongly}. Experimentally verifying the validity of this model for the DKS is a prerequisite for exploring its application in quantum technologies. 

\begin{figure*}[t!]
\centering
\includegraphics[width=\linewidth]{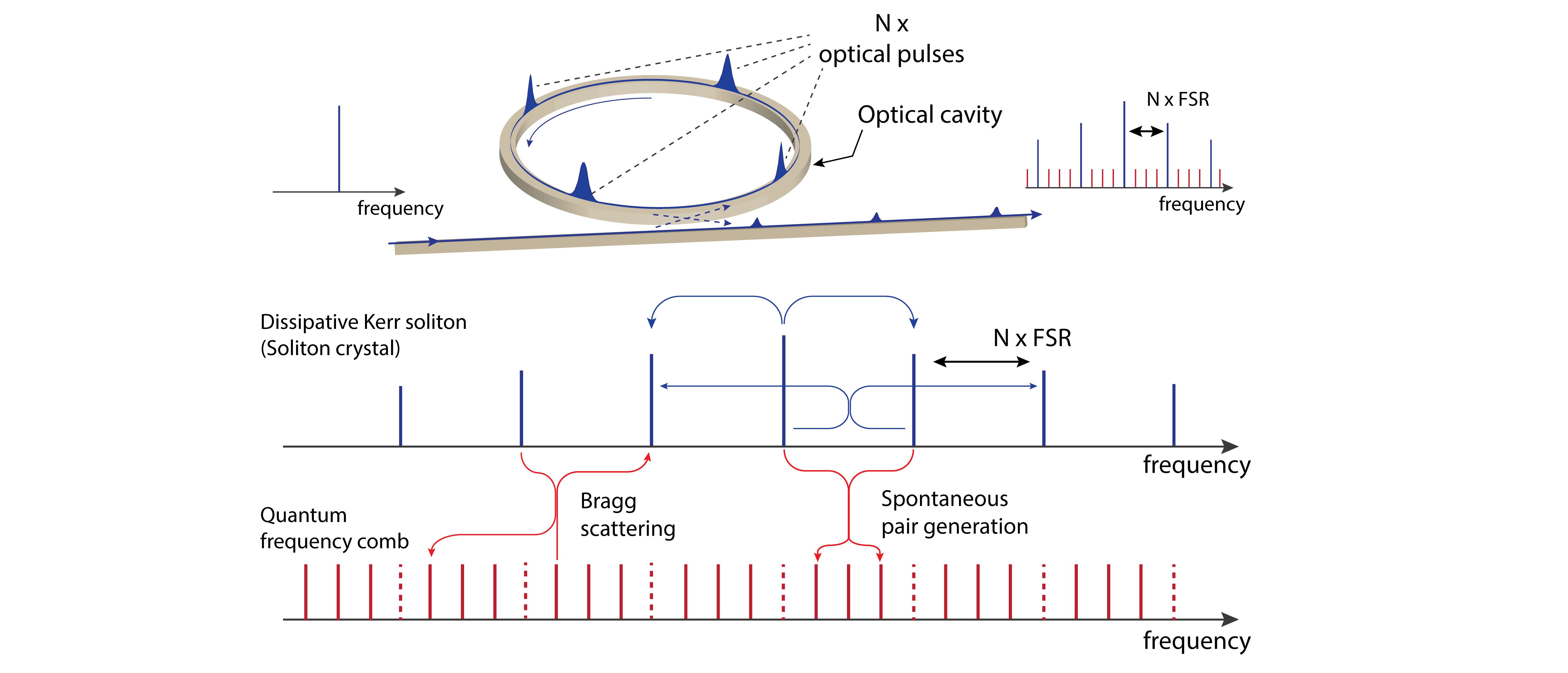}
\captionsetup{format=plain,justification=RaggedRight}
\caption{{\bf{Linearized model for quantum optical fields in a DKS state}} A schematic depiction of a Kerr microresonator with a circulating perfect soliton crystal state. The full optical state is modeled as a coherent classical comb (blue) that drives the quantum comb (red) via spontaneous parametric processes.}
\label{fig:model}
\end{figure*}

In this work, we use second-order photon correlation measurements to experimentally validate the linearized model (Eq.~\ref{eq:H}) for describing the spontaneous parametric processes in DKS states. We study the quantum formation dynamics of Kerr frequency combs and identify a class of DKS states --- perfect soliton crystals\cite{Papp:2017:NatPhoton, Kippenberg:2019:SolitonCrystals} --- which naturally isolate a subset of quantum optical fields from the coherent mean-field (Fig.~\ref{fig:model}). We measure the correlation matrix of the below-threshold modes to numerically infer the multi-mode entanglement structure of the state. We find that the DKS state may be engineered as an on-chip source of all-to-all entanglement.

% Figure 1 

\begin{figure*}[t!]
\centering
\includegraphics[width=\linewidth]{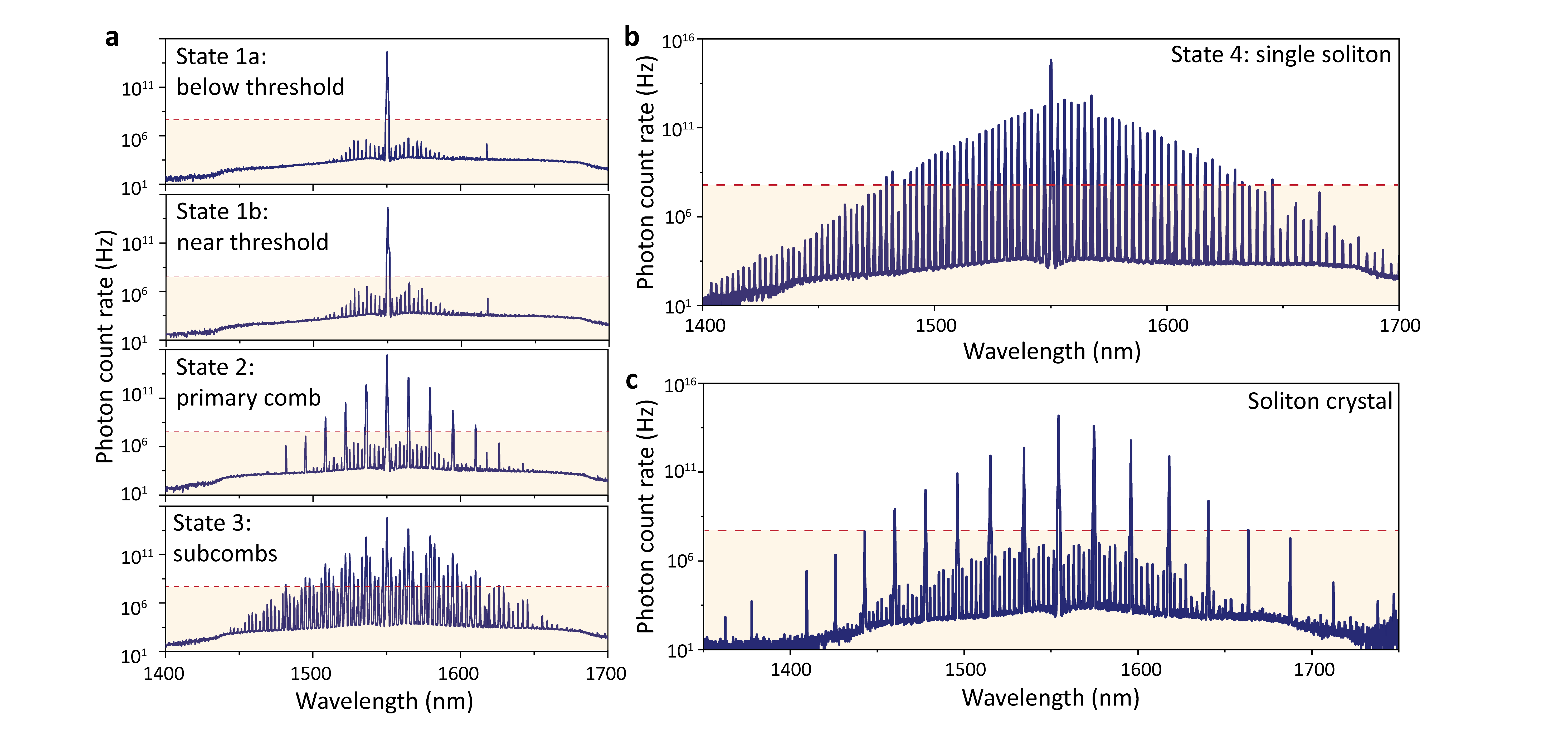}
\captionsetup{format=plain,justification=RaggedRight}
\caption{{\bf{Single-photon spectroscopy of optical microcombs}}  \textbf{(a)} Stages of frequency comb formation observed on the single-photon optical spectrum analyzer (SPOSA). Dashed line indicates noise floor of a commercial optical spectrum analyzer (-80~dBm). \textbf{(b)} The single soliton state. \textbf{(c)} A 7-FSR soliton crystal state, observed in a different device.}
\label{fig:setup}
\end{figure*}

% Figure 2 
\begin{figure*}[t!]
\centering
\includegraphics[width=\linewidth]{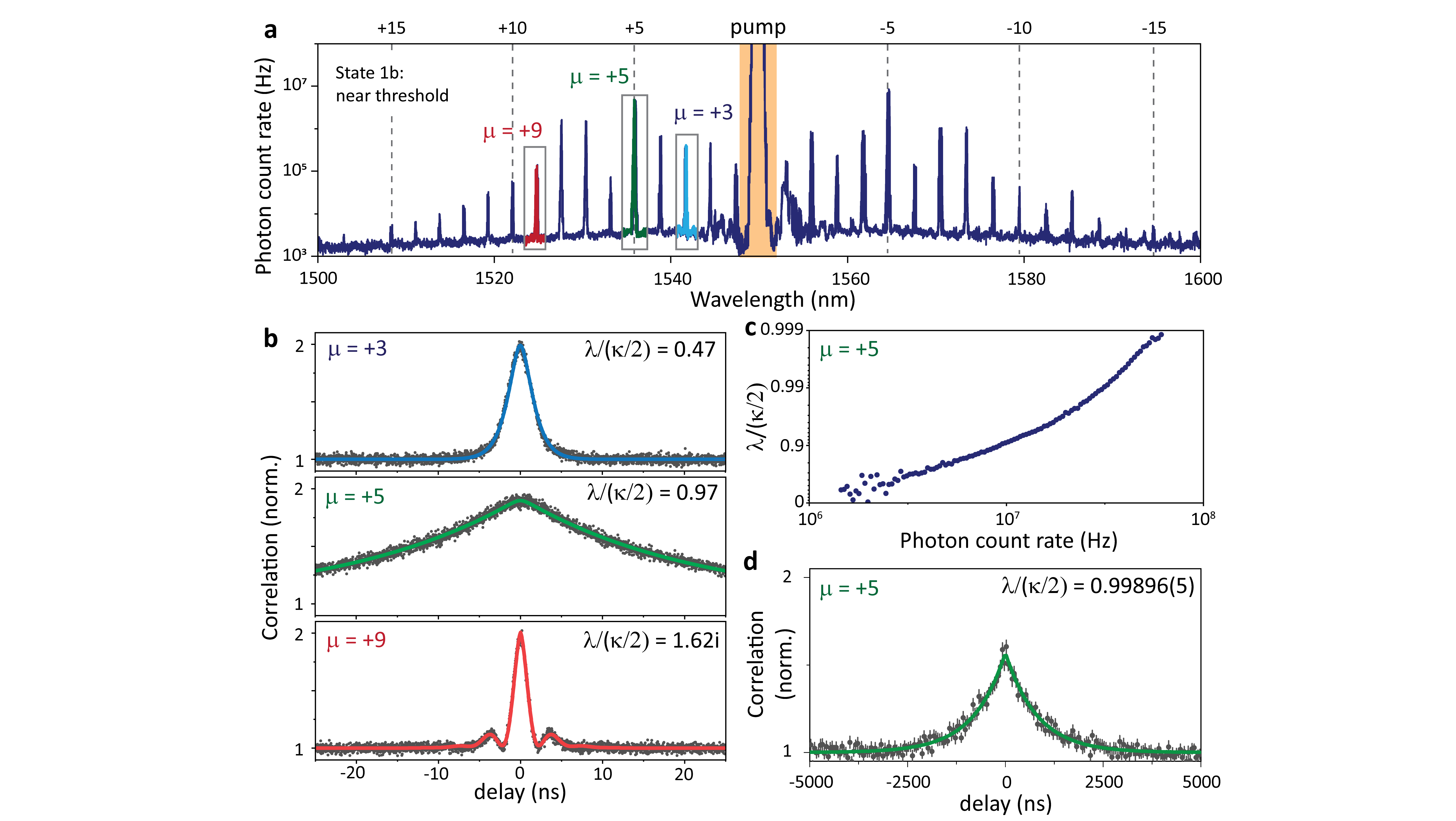}
\captionsetup{format=plain,justification=RaggedRight}
\caption{{\bf{Quantum coherence of parametric oscillation}} \textbf{(a)} The near-threshold spectrum reproduced from Fig.~\ref{fig:setup}a. Vertical dashed lines indicate the modes where the primary comb will form. \textbf{(b)} Measured $g^{(2)}_{\text{auto}}(\tau)$ on different modes shows the dispersion-dependent parametric gain variation throughout the comb. \textbf{(c)} Observation of asymptotic growth of coherence near the OPO threshold. The effective parametric gain $\lambda$, extracted from a numerical fit to Eq.~\ref{eq:auto} is plotted against the detected count rate on mode ${\mu=+5}$. \textbf{(d)} At the highest photon count rate, $g^{(2)}_{\text{auto}}(\tau)$ reveals coherence-broadening which corresponds to threshold proximity of 0.99896(5).}
\label{fig:threshold}
\end{figure*}

% Figure 3 
\begin{figure*}[t!]
\centering
\includegraphics[width=\linewidth]{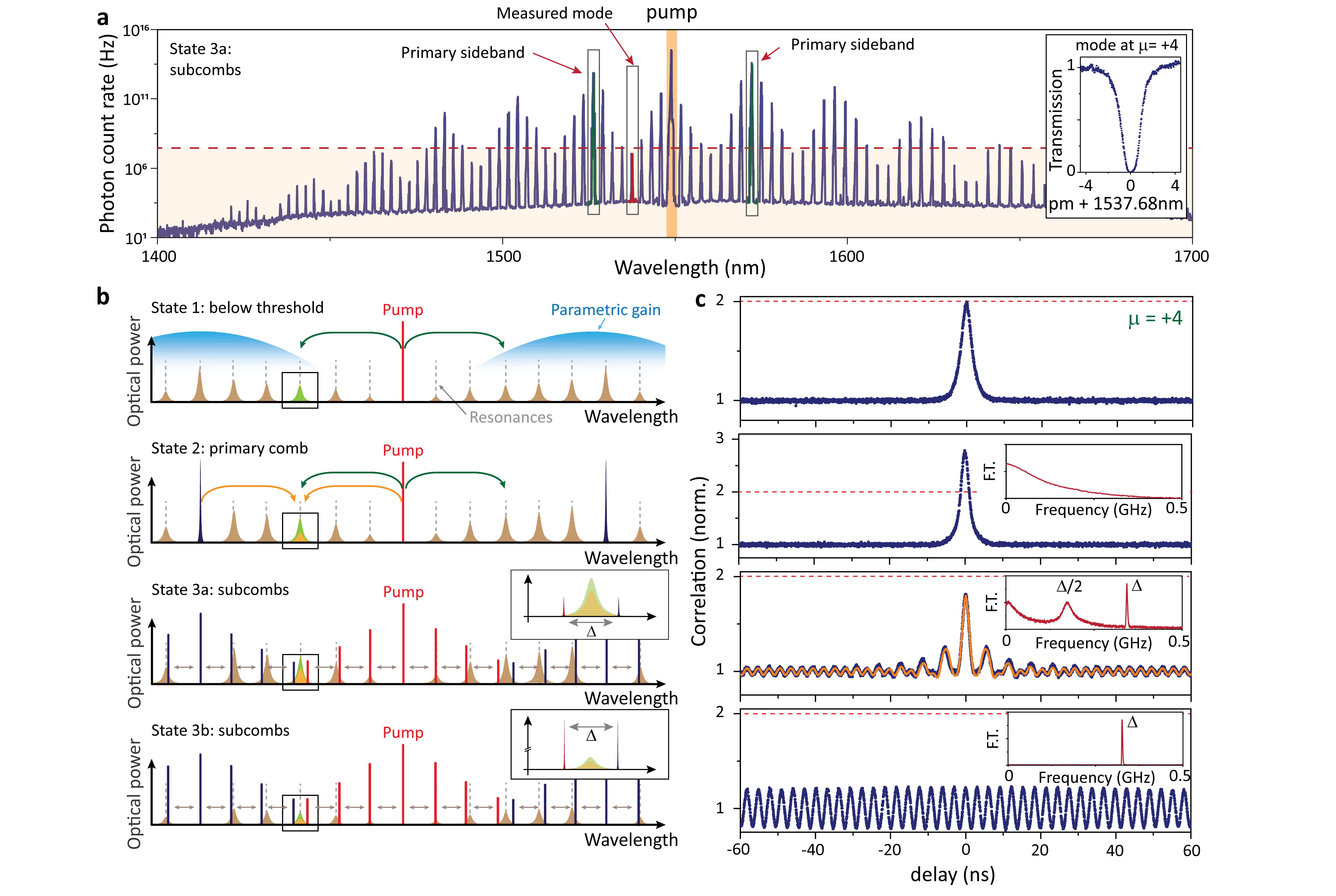}
\captionsetup{format=plain,justification=RaggedRight}
\caption{{\bf{Formation dynamics of secondary combs}} \textbf{(a)} A SPOSA spectrum of a secondary comb state.  \textbf{(b)} A graphical representation of the stages of secondary comb formation. \textbf{(c)} The auto-correlation is measured at mode ${\mu = +4}$ for each comb state. In state 1, only non-degenerate pair generation contributes to mode ${\mu=+4}$. In state 2, simultaneous degenerate and non-degenerate spontaneous pair generation is present. In state 3a, subcombs begin to merge and two-photon correlations reveal the interference of quantum fluctuations with the coherent state. The data are overlaid with the fit to input-output theory. In state 3b, the coherent state dominates and only the RF beat note is observed.}
\label{fig:secondary}
\end{figure*}

The observation of the full optical spectrum of a Kerr frequency comb, including the above- and below-threshold processes, requires single-photon sensitivity and a high dynamic range. For this purpose, we designed a single-photon optical spectrum analyzer (SPOSA) using multipass grating monochromators and superconducting nanowire single-photon detectors (SNSPDs) (PhotonSpot, Inc). The SPOSA has a broadband (${>200}$~nm) quantum efficiency of $\sim20\%$, and close-in dynamic range of $>140$~dB at $\pm0.8$~nm. We generate Kerr frequency combs using microring resonators with 350~GHz free spectral range (FSR) fabricated in 4H-silicon~carbide-on-insulator, a platform with favorable nonlinear\cite{guidry2020optical} and quantum \cite{lukin20204h} optical properties. With intrinsic quality factors (Q) as high as $5.6\times10^6$, low threshold OPO and low-power soliton operation (0.5~mW and 2.3~mW, respectively) are achieved. We use the SPOSA to characterize the Kerr microcomb spectra through all stages of DKS formation (Fig.~\ref{fig:setup}a,b). The $\sech^2$ soliton envelope is seen to persist at the tails of the DKS, in modes with very low photon number $\expval{a^\dagger_ja_j}<10^{-3}$. The quantum fluctuations generated in the DKS state and responsible for its quantum-limited timing jitter \cite{matsko2013timing, bao2021quantum} are obscured in the single-soliton spectrum. A perfect soliton crystal \cite{Papp:2017:NatPhoton, Kippenberg:2019:SolitonCrystals} (henceforth referred to as a soliton crystal), however, reveals these quantum fluctuations (Fig.~\ref{fig:setup}c). 

The formation of DKS from a below-threshold quantum frequency comb begins with the transition of a spontaneous FWM process into a stimulated FWM process at the onset of OPO. This transition can be observed through the second-order correlation function, $g^{(2)}(\tau)$. Using input-output theory\cite{ou1999cavity} we derive the exact form of $g^{(2)}(\tau)$ for a general two-mode parametric process. The general derivation is presented in the Supplementary Information. For signal and idler modes of equal linewidths $\kappa$, the auto-correlation is given by
\begin{equation}
g_{\text{auto}}^{(2)}(\tau) = 1 + \frac{e^{-\kappa  \tau}}{\lambda^2} \Big[ \frac{\kappa}{2}  \sinh (\lambda  \tau)+ \lambda  \cosh ( \lambda \tau) \Big]^2
\label{eq:auto}
\end{equation}
where $\lambda = \sqrt{ g^2 - \delta^2}$ is the effective parametric gain, $g = g_0 |A_0|^2$ is the mode coupling strength, and $\delta$ is the detuning.  Here, two regimes are notable: when $\delta^2>g^2$, $\lambda$ is imaginary which gives rise to oscillations in $g_{\text{auto}}^{(2)}(\tau)$, corresponding to the double-peaked photon spectrum for a strongly detuned parametric process\cite{chembo2016quantum, vernon2015strongly}; when $\lambda$ approaches $\kappa/2$, the $g_{\text{auto}}^{(2)}(\tau)$ coherence (decay time) increases asymptotically, reflecting the transition of the spontaneous FWM process into a stimulated FWM process (analogous to lasing). In a Kerr frequency comb close to but below the OPO threshold (Fig.~\ref{fig:threshold}a), both regimes can be observed simultaneously. We measure $g^{(2)}_{\text{auto}}(\tau)$ using two SNSPD detectors in a Hanbury Brown and Twiss configuration, and observe the dispersion-induced variation in $\lambda$ for different signal-idler pairs (Fig.~\ref{fig:threshold}b).\cite{herr2012universal}. Mode $\mu=+9$, far away from the pump, displays oscillations in  $g^{(2)}_{\text{auto}}(\tau)$, signifying poor phase-matching. In contrast, mode ${\mu=+5}$ shows a substantial coherence increase, which correctly predicts that it will seed the formation of the primary comb. To observe the asymptotic coherence growth at threshold, we repeatedly sweep the pump laser detuning through the OPO threshold condition, while synchronously acquiring the photon count rates and two-photon correlations. Through the numerical fit to Eq.~\ref{eq:auto}, $\lambda$ is extracted and plotted against the measured count rate (Fig.~\ref{fig:threshold}c). The maximum recorded coherence broadening exceeds the cavity coherence\cite{ou1999cavity} by nearly three orders of magnitude (Fig.~\ref{fig:threshold}d), which indicates that the state is approaching the critical point at which the linearized model would break down.\cite{navarrete2014regularized, vernon2015strongly}

% Figure 4
\begin{figure*}[t!]
\centering
\includegraphics[width=\linewidth]{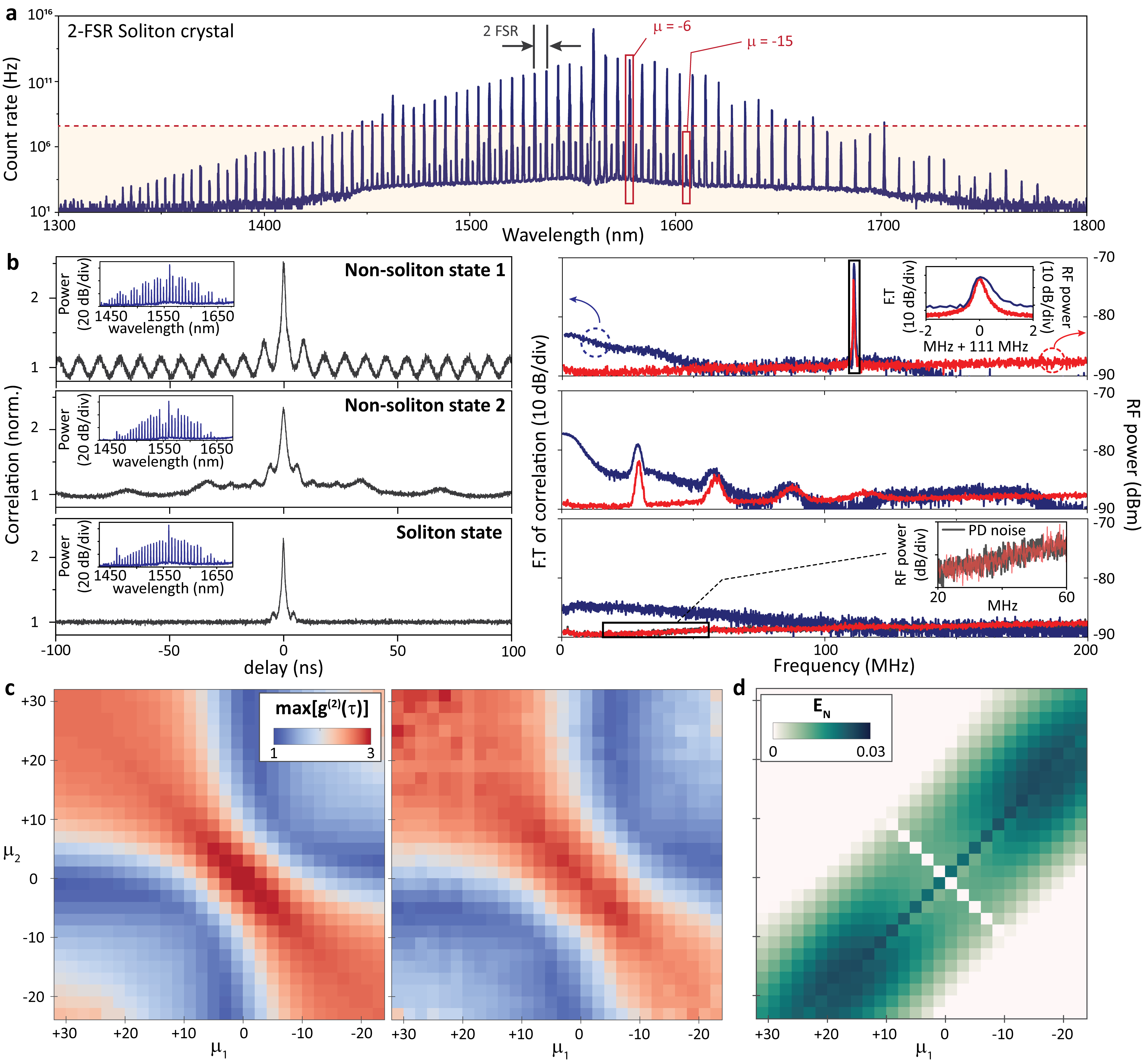}
\captionsetup{format=plain,justification=RaggedRight}
\caption{{\bf{Quantum correlations in non-phase-locked combs and perfect soliton crystals}} \textbf{(a)}A SPOSA spectrum of a 2-FSR soliton crystal. \textbf{(b)} Left: The measured $g^{(2)}(\tau)$ at mode ${\mu=-15}$ for different 2-FSR states observed in the device (inset shows an OSA spectrum of the measured state). Right: The RF beat note measured on a photodetector at mode ${\mu=-6}$ (red), and the Fourier transform of the measured $g^{(2)}(\tau)$ (blue). Bottom panel includes the photodetector noise floor, corroborating the low-noise soliton state. \textbf{(c)} The second-order correlations matrix for the below-threshold modes in the 2-FSR soliton crystal. Left: theoretical model; Right: experimental data. \textbf{(d)} The logarithmic negativity ($E_N$) matrix calculated for the 2-FSR soliton crystal assuming 10$\times$ increased out-coupling of the below-threshold modes. No pairwise entanglement is predicted in the device with unmodified waveguide coupling.}
\label{fig:soliton_crystal}
\end{figure*} 

In the linearized model of an above-threshold Kerr comb, a resonator mode may be occupied by both a coherent state and quantum fluctuations. In theory, the interference of the coherent state and the quantum fluctuations can be revealed through $g^{(2)}(\tau)$, but the intensity of the former is usually orders of magnitude greater (as can be seen in the spectra of the primary comb and the soliton crystal, Fig.~\ref{fig:setup}), making the experimental observation difficult. An exception can be found in the formation of secondary combs. We characterize the quantum formation dynamics of secondary combs through second-order photon correlations (Fig.~\ref{fig:secondary}), complementing earlier classical studies \cite{herr2012universal, chembo:2019:SecondaryCombs}. By monitoring mode $\mu=+4$, which is equidistant from the pump and the primary sideband, we observe simultaneously the degenerate and non-degenerate spontaneous pair generation processes, as well as the merging of the coherent subcombs. Notably, at the onset of subcombs merging, the intensities of the (bichromatic) coherent state and the quantum optical fields become comparable, and the signature of their interference in two-photon correlations is readily observed (Fig.~\ref{fig:secondary}c, state 3a). The Fourier transform of the auto-correlation shows three peaks, at 0, $\Delta/2$, and $\Delta$; they represent, respectively, the two-photon bunching of spontaneous pair generation, the interference of the quantum fluctuations with the bichromatic coherent state, and the coherent RF beat note\cite{herr2012universal, chembo:2019:SecondaryCombs} of the coherent state. The details of the theory used to model the interference are presented in the Supplementary Information. As the subcombs continue to merge, the coherent light drowns out the spontaneous parametric processes, and the photon correlations correspond to the interference of two weak coherent sources.\cite{da2015spectral} 

In a DKS state, the coherent comb is phase-locked and time-independent in the group-velocity reference frame, yielding a time-independent Hamiltonian for the quantum optical fields (Eq.~\ref{eq:H}). In contrast, a comb that is not phase-locked will produce a Hamiltonian with a time-dependent drive. To observe this effect in experiment, we consider a microring resonator which supports three distinct states with a 2-FSR spacing: one state is a 2-FSR soliton crystal (Fig.~\ref{fig:soliton_crystal}a); the others are non-natively spaced secondary combs\cite{herr2012universal} in the process of merging. These secondary combs are non-phase-locked states, which manifests in the frequency domain as polychromatic comb teeth. For each 2-FSR state, we measure $g^{(2)}(\tau) = \expval{g^{(2)}(t, \tau)}_t$ on mode ${\mu=-15}$ while simultaneously measuring the RF spectrum of mode ${\mu=-6}$ on a photodetector. Whereas the soliton crystal two-photon correlations are time-independent far from zero time delay, the correlations of a non-soliton state exhibit oscillations whose Fourier transform matches the RF spectrum measured on the photodetector. Such temporal dynamics may be modelled via Floquet theory\cite{navarrete2021floquet}.  We note that similar temporal oscillations in $g^{(2)}(\tau)$ have been observed in Floquet-driven two-level systems\cite{lukin2020spectrally}.

The soliton crystal state offers an excellent opportunity to experimentally verify the linearized model for the DKS state. The mean-field solution (complex amplitudes $A_m$ in Eq.~\ref{eq:H}) of the soliton crystal can be readily computed via the LLE; the below-threshold modes comprise the quantum fluctuations driven by the mean-field soliton without any admixture of coherent light and, crucially, they are decoupled from the the quantum fluctuations of the above-threshold modes by the mode matching condition $\delta_{\text{FWM}}$. We measure the correlations between all pairs of below-threshold modes of the 2-FSR soliton crystal, and compute the theoretically predicted second-order correlations matrix for the measured device parameters (Fig.~\ref{fig:soliton_crystal}c). We note that the only free parameter in the model is the pump laser detuning (within the soliton locking range). The agreement of the model with the experiment suggests that the quadratic Hamiltonian of the linearized model is indeed appropriate for describing the photon statistics of the quantum optical field generated in the below-threshold modes of a DKS state. 

We perform an analysis of the multipartite entanglement across the 2-FSR soliton crystal in the resonator mode basis by computing the logarithmic negativity\cite{vidal2002computable}, $E_N$, for all mode pairs. We find that in this basis, no pair-wise entanglement is predicted in the measured device. We then consider a modified device architecture, where the below-threshold modes are overcoupled to the output waveguide via a photonic molecule configuration (two coupled microresonators) \cite{zhang2019electronically, helgason2021dissipative}. This architecture is advantageous because it allows for the efficient extraction of the quantum optical fields from the device while simultaneously filtering them from the coherent fields, all without impacting the soliton crystal (see Supplementary Information). For this system, we numerically observe all-to-all entanglement along the signal-idler diagonal of the pump (Fig.~\ref{fig:soliton_crystal}d). Such entanglement structure is consistent with the all-to-all connectivity in the 2-FSR soliton crystal Hamiltonian. In ${(N>3)}$-FSR soliton crystals, the below threshold modes are not all-to-all coupled. Instead, the modes are divided into disjoint sets grouped by the value {${|\mu \mod N|}$}, as per the mode-matching condition. Thus, $\left \lfloor{N/2}\right \rfloor $ non-interacting subsets (each internally all-to-all coupled) are expected in an N-FSR soliton crystal. Indeed, for the 7-FSR soliton crystal state presented in Fig.~\ref{fig:setup}d, we experimentally confirm three disjoint all-to-all correlated sets of modes (see Supplementary Information).

In conclusion, we have investigated the quantum formation dynamics of DKS states and their quantum correlations. The experimental methods introduced here can serve as a starting point for exploring higher-order corrections to the linearized model both in the DKS and at near-threshold critical points. Applications of DKS states in quantum technologies will be informed by further theoretical and experimental characterization of the intra-comb entanglement. Once target supermodes\cite{roslund2014wavelength} and nullifiers\cite{cai2017multimode,ra2020non} are identified for the system, dispersion engineering\cite{Brasch:2016:Science, Lu:2019:NaturePhysics} can be employed to design the desired system Hamiltonian. The recent demonstrations of soliton generation in CMOS-foundry photonics\cite{jin2021hertz, xiang2021laser}, efficient on-chip frequency translation\cite{hu2020reconfigurable}, integrated detection of squeezed light\cite{tasker2021silicon}, and on-chip squeezed microcombs\cite{yang2021squeezed} lay out a clear path toward developing the DKS state as a resource for continuous variable quantum information processing \cite{xie2015harnessing, Furasawa:2019:ClusterState, Andresen:2019:ClusterState, zhong2020quantum}.

The large-scale multi-mode entanglement possible in a DKS state may also find applications in discrete variable quantum computation protocols operating under continuous wave (CW) \cite{Weiner:2017:Optica, Samara:2020:Swapping} or pulsed\cite{Morandotti:2016:Science, kues2017chip} drive. Although the soliton is traditionally operated in the CW regime, the recent study of solitons driven by a pulsed pump may enable the implementation of time-bin entanglement protocols\cite{daugey2021kerr} and the study of the temporal dynamics of quantum correlations\cite{imany2020probing} under Floquet drive. Pulsed operation can also be achieved with a CW-driven soliton via a time-dependent coupling constant $g_0(t)$ (Eq.~\ref{eq:H}), via the rapidly-tunable photonic molecule configuration\cite{zhang2019electronically}, to realize multimode control of spectral and temporal entanglement in a single integrated photonic device. Finally, we note that with the recent demonstration of Pockels soliton microcombs\cite{bruch2021pockels}, our results may be further extended to interactions between second- and third-order parametric processes.\\

$^*$These authors contributed equally.

$^\dagger$ jela{\makeatletter @\makeatother}stanford.edu \\

\noindent\textbf{Acknowledgments}
We gratefully acknowledge discussions with John Bowers, Tian Zhong, Lin Chang, Chengying Bao, Boqiang Shen and Avik Dutt. This work is funded by the Defense Advanced Research Projects Agency under the PIPES program. M.A.G. acknowledges the Albion Hewlett Stanford Graduate Fellowship (SGF) and the NSF Graduate Research Fellowship. D.M.L. acknowledges the Fong SGF and the National Defense Science and Engineering Graduate Fellowship. Part of this work was performed at the Stanford Nanofabrication Facility (SNF) and the Stanford Nano Shared Facilities (SNSF).

\bibliography{main}

\end{document}

\title{Supplementary Information for \\ Quantum Optics of Soliton Microcombs}
\author{Melissa A. Guidry$^{*,1}$, Daniil M. Lukin$^{*,1}$, Ki Youl Yang$^{*,1}$, Rahul Trivedi$^{1,2}$ and Jelena Vu\v{c}kovi\'{c}$^{\dagger,1}$\\
\vspace{+0.05 in}
$^1$E. L. Ginzton Laboratory, Stanford University, Stanford, CA 94305, USA.\\
$^2$Max-Planck-Institute for Quantum Optics, Hans-Kopfermann-Str. 1, 85748 Garching, Germany
}

%\appendix 
\renewcommand{\thefigure}{S\arabic{figure}}
\renewcommand{\thesection}{\Roman{section}}
\renewcommand{\bibnumfmt}[1]{[S#1]}
\renewcommand{\citenumfont}[1]{S#1}
\setcounter{figure}{0}
\setcounter{section}{0}

\maketitle

\onecolumngrid 

\tableofcontents

\newpage

\section{Linearized model of the soliton microcomb}
In this section, we detail the linearized model and the calculation of second-order photon correlations. First, we begin with a general quartic Hamiltonian for a Kerr resonator with a single pump; we linearize the model to produce a quadratic Hamiltonian, where modes are coupled through the classical amplitudes which can be described by the LLE\cite{chembo::2013::LLE}. Next, we define the input-output equations for the open quantum system, which we use to calculate the second-order correlation functions in a below-threshold comb. We numerically model the experiment of Fig.~3 from the main text to demonstrate the interface of the LLE with the input-output formalism.  We then extend our model to describe photon correlations in the presence of bichromatic coherent light. Finally, we describe how to characterize entanglement (the logarithmic negativity, $E_N$) between different cavity modes starting from the Heisenberg equations. 

\subsection{Linearization}
We consider the most general system Hamiltonian for four-wave mixing between cavity modes with coherent drive of the pump mode, $\mu = 0$:
\begin{align}
    \hat{H}_{\text{sys}} = \sum_\mu \omega_\mu \hat{a}_{\mu}^\dag \hat{a}_{\mu}  -\frac{1}{2} g_0 \sum_{\mu,\nu,j,k} \delta[ \mu + \nu - j - k ] \hat{a}_{\mu}^\dag \hat{a}_{\nu}^\dag \hat{a}_{j}  \hat{a}_{k} + \alpha_0 ( \hat{a}_0 e^{i \omega_{p} t} + \hat{a}_0^\dag e^{-i \omega_{p} t } ) 
\end{align}
where $\delta[ \mu + \nu - j - k ]$ is the Kronecker delta which enforces the four-wave mixing mode-matching condition. Here, $\omega_\mu$ is the resonance frequency of cavity mode $\mu$ and $\omega_{p}$ is the frequency of the coherent pump driving the central mode. The nonlinear coupling coefficient
$$ g_0 = \frac{ \hbar \omega_{p}^2 c n_2  }{n_0^2 V_{\text{eff}}} $$
represents the per photon frequency shift of the cavity due to the third-order nonlinearity of the cavity: $\hbar$ is the reduced Planck's constant, $n_2$ is the nonlinear refractive index, $n_0$ is the material index, and $V_{\text{eff}}$ is the effective mode volume of the resonator. The amplitude of the drive field is
$$ \alpha_0 =  \sqrt{\frac{\kappa_{c} P_{\text{wg}}}{\hbar \omega_{p}}} $$
where $\kappa_c$ is the coupling rate of the cavity to the input waveguide and $P_{\text{wg}}$ is the power in the input waveguide. To linearize the system, we write a formal separation of the optical state into the mean field solution (assumed to be a coherent state) and the quantum fluctuations: $\hat{a}_\mu(t) \rightarrow \alpha_\mu(t) + \hat{a}_\mu(t)$, where $\alpha_\mu(t)$ is the complex amplitude of the coherent state inside the cavity.
Moving into the reference frame which removes explicit time dependence from the classical coupled mode equations\cite{chembo2016quantum}, we apply a unitary transformation using $\hat{U}(t) = e^{i \hat{R} t}$ where $\hat{R} = \sum_\mu (\omega_{p} + D_1 \mu ) \hat{a}_{\mu}^\dag \hat{a}_{\mu}$. We enter the rotating frame of an evenly-spaced frequency ruler, with spacing $D_1$, centered at the central pump mode. Defining $\delta_\mu = \omega_\mu - \omega_{p} - D_1 \mu$, and keeping only quadratic terms, we arrive at a Hamiltonian which includes the interaction Hamiltonian Eq.~(1) of the main text:
\begin{align}
    \hat{H}_{\text{sys}} = &\sum_\mu \delta_\mu \hat{a}_{\mu}^\dag \hat{a}_{\mu}   -\frac{g_0}{2} \sum_{\mu,\nu,j,k} \delta[\mu + \nu - j - k] ( \overbrace{A_\mu A_\nu \hat{a}_{j}^\dag \hat{a}_{k}^\dag}^{\mathclap{\text{spontaneous pair generation}}}  + \underbrace{A_k^* A_\nu \hat{a}_{j}^\dag \hat{a}_{\mu}}_{\mathclap{\text{XPM \& Bragg scattering}}} + \text{h.c.} )
\end{align}
where $A_\mu$ are the complex-valued field amplitudes, described using the Lugiato-Lefever equation\cite{chembo::2013::LLE}. The last term describes cross-phase modulation (XPM) when $\mu = j$, and four-wave mixing Bragg scattering otherwise.

\subsection{Input-output formalism}
The following calculations are in the Heisenberg picture. To describe the open system, we allow our cavity modes to couple to a bath. For a resonator with a large finesse, a Markovian approximation can be made on each mode independently
\begin{align}
    H_{\text{bath}} =  \sum_\mu \int \omega \hat{b}_\mu^\dagger(\omega) \hat{b}_\mu(\omega) d\omega
\end{align}
with a bath-coupling Hamiltonian
\begin{align}
    V = \sum_\mu \sqrt{\frac{\kappa_{\mu}}{2\pi}}\int \hat{a}_\mu^\dagger(\omega) \hat{b}_\mu(\omega) d\omega 
\end{align}
where $\kappa_\mu$ is the total loss rate for cavity mode $\mu$. Starting from the Hamiltonian $H = H_{\text{sys}} + H_{\text{bath}} + V$, we use the Heisenberg equations of motion $\dot{\hat{a}}_\mu = -i[\hat{a}_\mu,\hat{H}]$ to write down quantum coupled mode equations for this system which resemble the classical coupled mode equations:
\begin{align}
    \frac{d\hat{a}_\mu(t)}{dt} = & - \bigg(i\delta_\mu + \frac{\kappa_\mu}{2}\bigg)\hat{a}_\mu(t) + i g_0 \sum_{\nu,j,k} \delta[ \mu + \nu - j -k ]  A_j A_k \hat{a}_\nu^\dagger(t) \nonumber\\
    & + 2 i g_0 \sum_{\nu,j,k} \delta[ \mu + j - \nu -k ]  A_j^* A_k \hat{a}_\nu(t) - i\sqrt{\kappa_\mu} \hat{b}_{\text{in}, \mu}(t) 
\end{align}
Each bath-cavity mode pair has an associated input-output relation:
\begin{align}
    \hat{b}_{\text{out}, \mu}(t) = \hat{b}_{\text{in}, \mu}(t) - i\sqrt{\kappa_\mu} \hat{a}_\mu(t)
\end{align}
We define the following $2n$-dimensional vectors describing $n$ quantum modes in the frequency domain:
\begin{align}
    \bar{a}(\omega) = \begin{bmatrix}
    \hat{a}_1(\omega)\\ \vdots \\  \hat{a}_n(\omega)\\  \hat{a}^\dag_1(-\omega)\\ \vdots \\ \hat{a}^\dag_n(-\omega)\\
    \end{bmatrix}
    \hspace{2em}
    \bar{b}_{\text{in}}(\omega) = \begin{bmatrix}
    \hat{b}_{\text{in},1}(\omega)\\ \vdots \\  \hat{b}_{\text{in},n}(\omega)\\  \hat{b}^\dag_{\text{in},1}(-\omega)\\ \vdots \\ \hat{b}^\dag_{\text{in},n}(-\omega)\\
    \end{bmatrix}
    \hspace{2em}
    \bar{b}_{\text{out}}(\omega) = \begin{bmatrix}
    \hat{b}_{\text{out},1}(\omega)\\ \vdots \\  \hat{b}_{\text{out},n}(\omega)\\  \hat{b}^\dag_{\text{out},1}(-\omega)\\ \vdots \\ \hat{b}^\dag_{\text{out},n}(-\omega)\\
    \end{bmatrix}
\end{align}
We can define a matrix $N(\omega)$ from our coupled mode equations relating the output fluctuations to the input fluctuations:
\begin{align}
\bar{b}_{\text{out}}(\omega) = N(\omega) \bar{b}_{\text{in}}(\omega)
\end{align}

\subsection{Second-order photon correlations below threshold}
We derive analytic expressions for two-photon correlations between quantum modes which interact with a single pumped mode\cite{ou1999cavity}. In this case, our Hamiltonian is simple, and we only consider interactions between pairs of modes ($-\mu$, $+\mu$) centered around the pump. For this analysis, we choose $D_1 = (\omega_+ - \omega_-)/2\mu$ for which $\delta_+ = \delta_- = \delta$.
\begin{align}
    \hat{H}_{\text{sys}} &= \delta (\hat{a}_{-}^\dag \hat{a}_{-} +  \hat{a}_{+}^\dag \hat{a}_{+}   ) + i g ( \hat{a}_{-} \hat{a}_{+} -  \hat{a}_{-}^\dag \hat{a}_{+}^\dag  )
\end{align}
where $g = g_0 |A_0|^2$, which includes both the Kerr coupling strength and the intensity in the pumped cavity mode. Assuming $\kappa_+ = \kappa_- = \kappa$, the Heisenberg equations read
\begin{align}
    &\frac{d\hat{a}_{\pm}(t)}{dt} = -\bigg(i\delta + \frac{\kappa}{2}\bigg)\hat{a}_{\pm}(t) - g \hat{a}_{\mp}^\dagger(t) - i\sqrt{\kappa} \hat{b}_{\text{in}, \pm}(t) 
\end{align}
which gives us the following matrix $N(\omega)$:
$$N(\omega) = \left(
\begin{array}{cccc}
 1 +\frac{ \kappa  ( \kappa/2 - i (\delta + \omega) )}{ (g^2 - \delta^2) - (\kappa/2 - i
   \omega )^2} & 0 & 0 & \frac{ g \kappa }{(g^2 - \delta^2)-(\kappa/2 - i \omega )^2}
   \\
 0 & 1 +\frac{ \kappa  ( \kappa/2 - i (\delta + \omega) )}{ (g^2 - \delta^2) - (\kappa/2 - i
   \omega )^2} & \frac{ g \kappa }{(g^2 - \delta^2)-(\kappa/2 - i \omega )^2} & 0 \\
 0 & \frac{ g \kappa }{(g^2 - \delta^2)-(\kappa/2 - i \omega )^2} & 1 +\frac{ \kappa  ( \kappa/2 + i (\delta - \omega) )}{ (g^2 - \delta^2) - (\kappa/2 - i
   \omega )^2} & 0 \\
 \frac{ g \kappa }{(g^2 - \delta^2)-(\kappa/2 - i \omega )^2} & 0 & 0 & 1 +\frac{ \kappa  ( \kappa/2 + i (\delta - \omega) )}{ (g^2 - \delta^2) - (\kappa/2 - i
   \omega )^2}
   \\
\end{array}
\right)
$$
We calculate the two-photon correlation function:
\begin{align}
g^{(2)}_{ij}(t+\tau, t) =\frac{G^{(2)}(t+\tau, t)} {\expval{\hat{b}_{\text{out,i}}^\dagger(t) \hat{b}_{\text{out,i}}(t)} \expval{\hat{b}_{\text{out,j}}^\dagger(t+\tau) \hat{b}_{\text{out,j}}(t+\tau)} }  
\end{align}
where 
\begin{align}
G^{(2)}(t+\tau, t) =  \expval{ \hat{b}_{\text{out,i}}^\dagger(t) \hat{b}_{\text{out,j}}^\dagger(t + \tau) \hat{b}_{\text{out,j}}(t + \tau) \hat{b}_{\text{out,i}}(t)}.
\end{align}
The expectation value is taken with respect to the initial vacuum state. The time-domain output bath operator is related to the frequency-domain input bath operator via:
\begin{align}
    &\hat{b}_{\text{out,i}}(t) = \frac{1}{\sqrt{2\pi}} \int_{-\infty}^\infty d\omega e^{-i\omega t} \hat{b}_{\text{out,i}} (\omega) = \frac{1}{\sqrt{2\pi}} \int_{-\infty}^\infty d\omega e^{-i\omega t} \sum_{k=1}^n \Big( N_{ik}(\omega) \hat{b}_{\text{in,k}} (\omega) + N_{i(k+n)}(\omega) \hat{b}_{\text{in,k}}^\dag (-\omega) \Big)
\end{align}
From here, $g^{(2)}_{ij}(t+\tau, t)$ may be calculated using the commutation relations of the frequency-domain input bath operators:
$$\expval{\hat{b}_{\text{in,i}} (\omega) \hat{b}_{\text{in,j}}^\dag (-\omega')} = \delta_{ij} \delta(\omega + \omega') \hspace{2.5em} \expval{ \hat{b}_{\text{in,i}}^\dag (-\omega) \hat{b}_{\text{in,j}} (\omega')} = 0$$
After applying the commutation relations, one arrives at an expression where each term is a product of Fourier transforms. Defining $\lambda = \sqrt{g^2 - \delta^2}$, the result is:
\begin{align}
    g_{++}^{(2)}(\tau) &= 1 + \frac{e^{-\kappa  \tau}}{\lambda^2} \Big[ \frac{\kappa}{2}  \sinh (\lambda  \tau)+ \lambda  \cosh ( \lambda \tau) \Big]^2 \\
    g_{+-}^{(2)}(\tau) &= 1 + \frac{e^{-\kappa  \tau}}{g^2} \Big| \Big(\lambda - i  \frac{\kappa}{2} \frac{\delta}{\lambda} \Big) \sinh (\lambda \tau)+ \Big(\frac{\kappa}{2} - i \delta \Big) \cosh( \lambda  \tau) \Big|^2
\end{align}
and $g_{++}^{(2)}(\tau) = g_{--}^{(2)}(\tau)$, $g_{+-}^{(2)}(\tau) = g_{-+}^{(2)}(-\tau)$. 

\subsection{Numerical analysis of OPO threshold}
\begin{figure}[h!]
\centering
\includegraphics[width=0.67\linewidth]{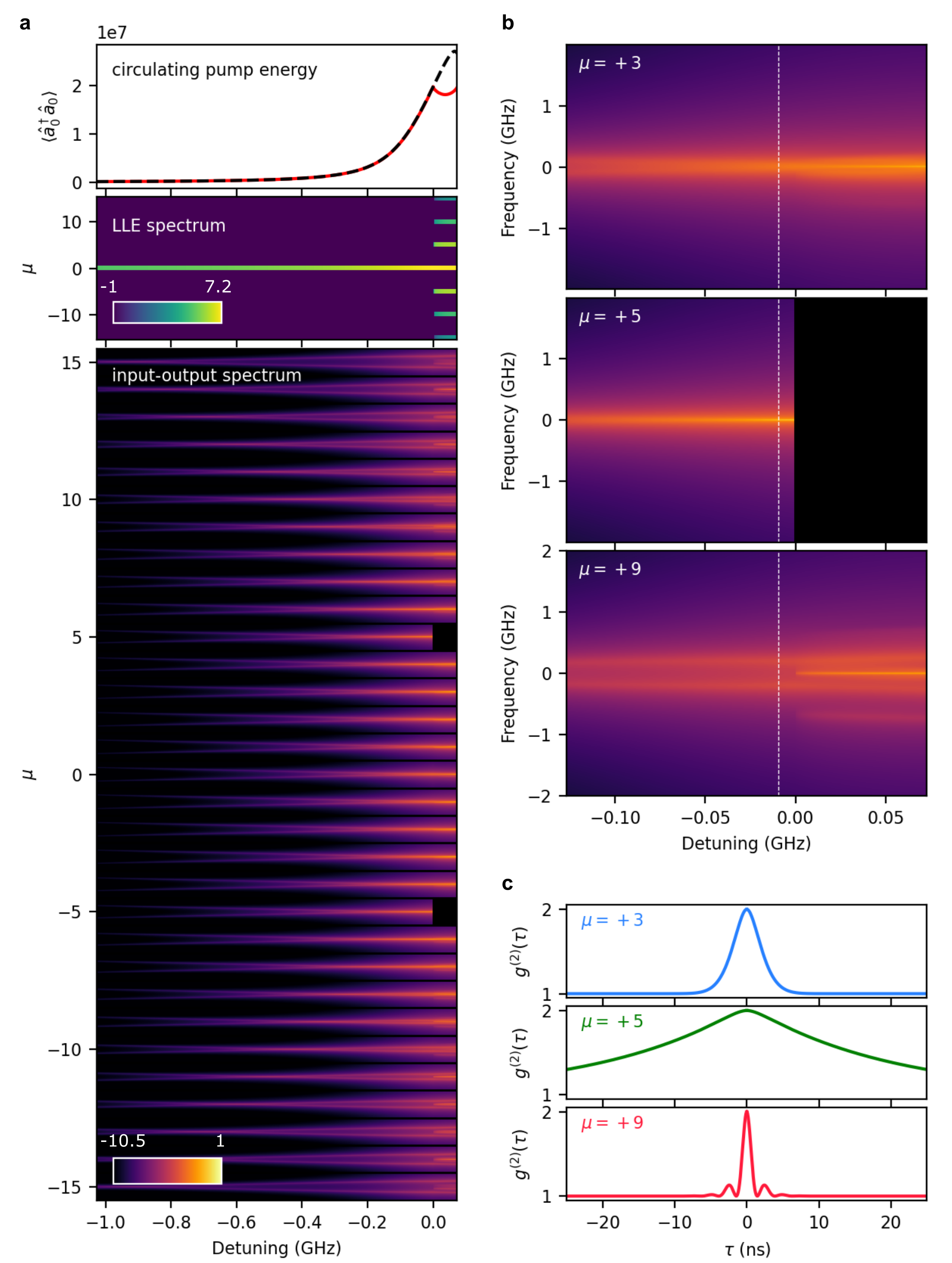}
\captionsetup{format=plain,justification=RaggedRight}
\caption{{\bf{Numerical analysis of OPO threshold}} \textbf{a} \textit{Top:} Dependence of the pump intensity in the microring (in units of photon number) on laser detuning. Detuning is given relative to the OPO threshold; a higher laser frequency corresponds to a more negative detuning with respect to the OPO condition. The LLE simulation (red) matches the analytic expression for the cavity mode in the presence of the Kerr nonlinear resonance shift (dashed black line) up to the threshold point, where the pump mode becomes depleted. \textit{Middle:} Spectrum of the classical Kerr comb in the ring resonator, computed via the LLE, showing the formation of the 5-FSR primary comb. The scale represents $\log_{10}(\text{\# of photons})$.  \textit{Bottom:} The photon spectrum of the below-threshold modes computed via input-output theory. For each mode, a spectral window of ${\pm2}$~GHz is shown. At threshold, modes ${\mu=\pm5}$ exit the regime of validity of the linearized model and are excluded from the input-output simulation.  The scale represents photon number spectral density, $\log_{10}(\text{\# of photons per Mrad/second})$. \textbf{b} Evolution of the spectrum of three select modes near threshold (same scale as \textit{(b)}). Above threshold, the spectra exhibit additional features generated by the additional parametric processes driven by the primary comb lines. \textbf{c} Computed $g^{(2)}_{\text{auto}}(\tau)$ at a detuning of $-9.3$~MHz (indicated as a dashed line in \textit{(b)}).
}
\label{fig:opo_threshold_LLE}
\end{figure}

To compute the second-order correlation matrix of the quantum optical fields in the presence of the above-threshold Kerr comb, we combine the LLE simulation with input-output theory\cite{chembo2016quantum}. The basic test case for the self-consistency of the joint LLE and input-output modelling is the OPO threshold condition: specifically, when the laser is tuned from blue to red to model the experiment, the asymptotic bandwidth narrowing in the below-threshold mode (as computed via input-output theory) and the formation of primary combs (as computed via the LLE) should happen simultaneously. We numerically reproduce the near-threshold behavior presented in Fig.~3 of the main text. The result of the combined LLE and input-output theory simulation is presented in Fig.~\ref{fig:opo_threshold_LLE}a. The modes ${\mu=\pm5}$ indeed exit the regime of validity of the input-output formalism (when $Q_{\text{eff}}$ has positive real eigenvalues) at the onset of the LLE threshold. Although coherent comb light is present in other modes above threshold, only  ${\mu=\pm5}$ modes cannot be simulated with the presented linearization approach.
A qualitative agreement between simulation and experimentally measured correlations (shown in Fig.~3b of the main text) is seen at the simulated detuning of $-9.3$~MHz, Fig.~\ref{fig:opo_threshold_LLE}c.

\subsection{Second-order photon correlations in merging secondary combs}
In this section we describe the modelling of the interference in $g^{(2)}(\tau)$ of stimulated and spontaneous FWM presented in Fig.~4 of the main text. The operator $\hat{b}_{\text{out}}(t)$ is:
\begin{align}
    \hat{b}_{\text{out}}(t) = \hat{b}_{\text{in}}(t) - i \sqrt{\kappa} \big[\alpha(t) + \hat{a}(t)\big],
\end{align}
or, in the frequency domain, 
\begin{align}
\hat{b}_{\text{out}}(\omega) = N(\omega) \hat{b}_{\text{in}}(\omega) - i \sqrt{\kappa} \alpha(\omega).
\end{align}
Since the coherent field in the mode is bichromatic,  
\begin{align}
\alpha(t) = A_{\text{coh,1}} e^{-i\omega_1 t} + A_{\text{coh,2}} e^{-i\omega_2 t}.
\end{align}
We note that in the presence of a bichromatic coherent state, there is a distinction between $g^{(2)}(\tau)$ and the experimentally-measured correlations. Specifically, the measured correlations are the time-averaged two-photon coincidences $G^{(2)}(\tau) = \expval{G^{(2)}(t+\tau, t)}_t,$ normalized to the mean value at $\tau \to \infty$:
\begin{align}
g_{\text{exp}}^{(2)}(\tau) = \frac{\int_{-\infty}^{\infty}G^{(2)}(t+\tau, t) \dd{t}}{\lim_{T\to\infty}\int_T^\infty\int_{-\infty}^{\infty}G^{(2)}(t+\tau, t) \dd{t} \dd{\tau}}
\end{align}
To model the $g_{\text{exp}}^{(2)}(\tau)$ presented in Fig.~4 of the main text, we consider a system of three coherent drives ($A_{-8}, A_0, A_{+8}$), and four quantum modes ($\hat{a}_{-12}, \hat{a}_{-4}, \hat{a}_{4}, \hat{a}_{12}$). The effect of other coherent driving modes is assumed negligible, because their amplitude is much smaller. The fit parameters in the model are: 1) two pump amplitudes ($A_0$, and $A_{+8} = A_{-8}$); 2) two coherent state amplitudes ($ A_{\text{coh,1}}$ and $ A_{\text{coh,2}}$); 3) two coherent state frequencies ($\omega_1$ and $\omega_2$); and 4) the pump laser detuning $\delta_p$. The fit is presented in Fig.~4, state 3a of the main text.

\subsection{Characterizing entanglement between cavity modes in soliton crystals}
\noindent Denoting the modes under consideration by $a_\mu$ for $\mu \in \{1, 2 \dots N\}$ and assuming a quadratic Hamiltonian. In steady state ($t\to \infty$), the density matrix describing any two modes $\alpha$ and $\beta$ is described by a Gaussian Wigner function:
\begin{align}
    W_{\alpha, \beta}(q_{\alpha, \beta} = [x_\alpha, p_\alpha, x_\beta, p_\beta]^T) =  \frac{1}{\pi \sqrt{\text{Det}[\Sigma_{\alpha, \beta}]}} \exp\big({q_{\alpha, \beta}^T \Sigma_{\alpha, \beta} q_{\alpha, \beta}}\big)
\end{align} 
where $\Sigma_{\alpha, \beta} = \langle q_{\alpha, \beta}q_{\alpha, \beta}^T\rangle_W$ is the $4\times 4$ steady state correlation matrix formed from weyl ordered operators. The entanglement measure between these two modes can be computed as a log-negativity of this matrix, defined by
\begin{align}
    \mathcal{E}_{\alpha, \beta} = \text{max}[0, -\text{log}(\sqrt{2}\eta)]
\end{align}
where
\begin{align}   
    \eta = \sqrt{\Theta - \sqrt{\Theta^2 - 4\text{Det}(\Sigma_{\alpha, \beta})}}
\end{align}
and
\begin{align}
    \Theta = \text{Det}(\Sigma_{\alpha}) + \text{Det}(\Sigma_{\beta}) - 2\text{Det}(C)
\end{align}
and $\Sigma_{\alpha}, \Sigma_{\beta}$ and $C$ are defined as different blocks of $\Sigma_{\alpha, \beta}$
\begin{align}
    \Sigma_{\alpha, \beta} = \begin{bmatrix} \Sigma_{\alpha} & C \\ C^T & \Sigma_\beta
    \end{bmatrix}
\end{align}
To compute the correlation matrix $\Sigma_{\alpha, \beta}$ from a quadratic Hamiltonian, it is convenient to express the correlation elements in terms of annihilation operators. We can immediately note that
\begin{subequations}
\begin{align}
    &\langle x_{\alpha} x_\beta\rangle_W = \frac{1}{2}\bigg[\langle a_{\alpha}a_\beta\rangle + \langle a_\alpha^\dagger a_\beta^\dagger \rangle + \langle a_{\alpha}^\dagger a_\beta \rangle + \langle a_{\alpha} a_\beta^\dagger \rangle\bigg] \\
    &\langle p_{\alpha} p_{\beta} \rangle_W = -\frac{1}{2}\bigg[\langle a_{\alpha}a_\beta\rangle + \langle a_\alpha^\dagger a_\beta^\dagger \rangle - \langle a_{\alpha}^\dagger a_\beta \rangle - \langle a_{\alpha} a_\beta^\dagger \rangle\bigg]\\
    &\langle x_{\alpha}p_{\beta} \rangle_W = \langle p_{\beta}x_{\alpha} \rangle_W = \frac{i}{2}\bigg[\delta_{\alpha, \beta} + \langle a_{\alpha}^\dagger a_{\beta}^\dagger\rangle -\langle a_{\alpha} a_{\beta}\rangle + \langle a_{\beta}^\dagger a_{\alpha}\rangle - \langle a_\beta a_{\alpha}^\dagger\rangle\bigg]
\end{align}
\end{subequations}
The correlators for the annihilation operators required above can be easily calculated from the input-output formalism. Recall that for a quadratic, time-invariant Hamiltonian, the Heisenberg equations read
\begin{align}
    \frac{d}{dt} \begin{bmatrix} a(t) \\
    a^\dagger(t)
    \end{bmatrix} = Q_{\text{eff}} \begin{bmatrix} a(t) \\
    a^\dagger(t)
    \end{bmatrix} + M  \begin{bmatrix}
    b_{in}(t) \\
    b_{in}^\dagger(t)
    \end{bmatrix}
\end{align}
Physically, we expect eigenvalues of $Q_{\text{eff}}$ to all have negative real part so as to have a well defined steady state. We then obtain by integrating the above equations that as $t\to \infty$
\begin{align}
    \begin{bmatrix}
    a(t) \\
    a^\dagger(t)
    \end{bmatrix} = \int_0^t e^{Q_{\text{eff}}(t - \tau)}M \begin{bmatrix}
    b_{in}(\tau) \\
    b_{in}^\dagger(\tau)
    \end{bmatrix} d\tau = \int_0^t X e^{\Lambda (t - \tau)}X^{-1}M \begin{bmatrix}
    b_{in}(\tau) \\
    b_{in}^\dagger(\tau)
    \end{bmatrix}d\tau
\end{align}
where we can define the eigenvalue decomposition $Q_{\text{eff}} = X\Lambda X^{-1}$. It is now straightforward to calculate
\begin{align}
    \lim_{t\to \infty}\langle \begin{bmatrix} a(t) \\ a^\dagger(t) \end{bmatrix}\begin{bmatrix} a^\dagger(t) & a(t) \end{bmatrix}\rangle = \lim_{t\to\infty}\int_0^t X e^{\Lambda(t-\tau)}X^{-1}MM^\dagger X^{-\dagger} e^{\Lambda^*(t - \tau)} X^\dagger d\tau = X N X^\dagger
\end{align}
where $N$ is a matrix whose elements are given by
\begin{align}
N_{i, j} = \frac{\big[X^{-1}MJM^\dagger X^{-\dagger}\big]_{i, j}}{\lambda_i + \lambda_j^*}
\end{align}
defined in terms of the eigenvalues of $Q_{\text{eff}}$,  $\lambda_i$, where $J$ is a $2N \times 2N$ matrix:
\begin{align*}
J =
\left[\begin{array}{c| c }
  I(N)
  & 0 \\
  \hline
  0 &
  0
\end{array}\right]
 \end{align*}

\newpage 

\section{Experimental setup}
\begin{figure}[h!]
\centering
\includegraphics[width=\linewidth]{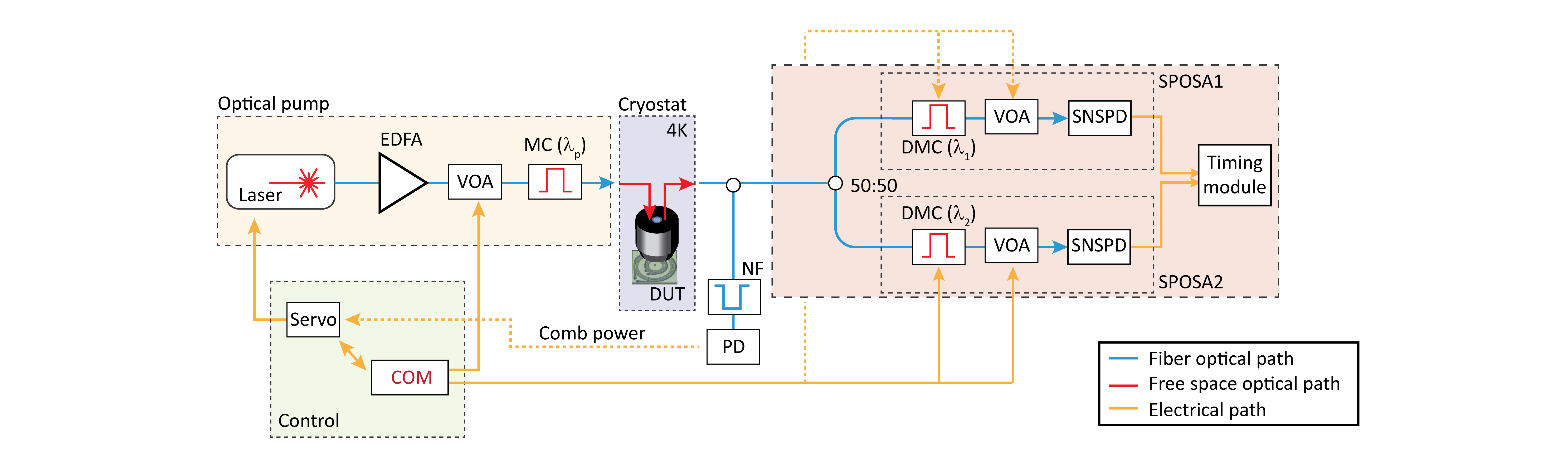}
\captionsetup{format=plain,justification=RaggedRight}
\caption{{\bf{Single-photon OSA}} Experimental setup. A silicon carbide microring resonator interfaced with inverse-designed vertical couplers\cite{guidry2020optical} is mounted in a closed-cycle cryostat. Automated locking of a desired microcomb state is performed using active feedback by controlling the laser wavelength and power. A free-space tunable two-pass monochromator (MC) and double monochromator (DMC) serve as narrow band-pass filters with rejection of $>$130~dB and $>$180~dB, respectively. The dynamic range of the superconducting nanowire single photon detectors (SNSPDs) is extended from 60~dB to 180~dB via variable optical attenuators (VOA). A single SPOSA is used for spectroscopy and two SPOSAs are used for photon correlation measurements. Photon detection events are recorded with a timing module TimeTagger Ultra from Swabian Instruments. }
\label{SPOSA}
\end{figure}

The experimental setup is presented in Fig.~\ref{SPOSA}. The device is operated at 4~K to reduce the thermo-optic response \cite{moille2019kerr}. In the design of the single-photon optical spectrum analyzer (SPOSA), we focused on broadband operation in order to image all parts of the frequency comb, which extends beyond the operation range of standard fiber-based filtering components. For this reason, a free-space monochromator approach was chosen. Blazed gratings (600 grooves/mm) optimized for 1.5~\textmu m operated in the Littrow configuration have $75-85$\% efficiency across the range of operation (1300-1700~nm). In a single-pass monochromator, dynamic range is limited to $\sim70$~dB by roughness-induced scattering from the grating surface. By operating the monochromator in a two-pass configuration, where the forward and return beams do not overlap on the grating, the dynamic range is doubled. To further increase the dynamic range, a second, single-pass monochromator can additionally be used. The double-pass monochromator configuration has dynamic range in excess of 180~dB and total peak efficiency as high as 55\%. The superconducting nanowire single photon detectors (SNSPDs) are optimized for broadband operation with efficiency exceeding 80\% from $1.0-1.6$~\textmu m.
\vspace{3em}

\newpage

\section{Silicon carbide soliton microcomb}
In this section, we describe the first demonstration of a soliton microcomb in a 4H-silicon carbide (SiC) microresonator. The CMOS-compatible fabrication process is described in \cite{lukin20204h, guidry2020optical}. SiC possesses a high linear and nonlinear refractive indices\cite{guidry2020optical} ($n=2.6$ and $n_2=6.9\cdot 10^{-15}$~cm\textsuperscript{2}/W at $1550$~nm), which makes it suitable for highly efficient, compact nonlinear photonic devices. However, the tight confinement and high material index of integrated waveguides make them susceptible to scattering losses caused by surface roughness. We demonstrate the fabrication of SiC microresonators with smooth sidewalls and strong confinement with record-high quality (Q) factors. The fabricated microring resonators have a radius of 100~\textmu m, height of 500-600~nm, and width of 1850~nm.  

\subsection{Sub-mW parametric oscillation threshold}
\begin{figure}[h!]
\centering
\includegraphics[width=0.9\linewidth]{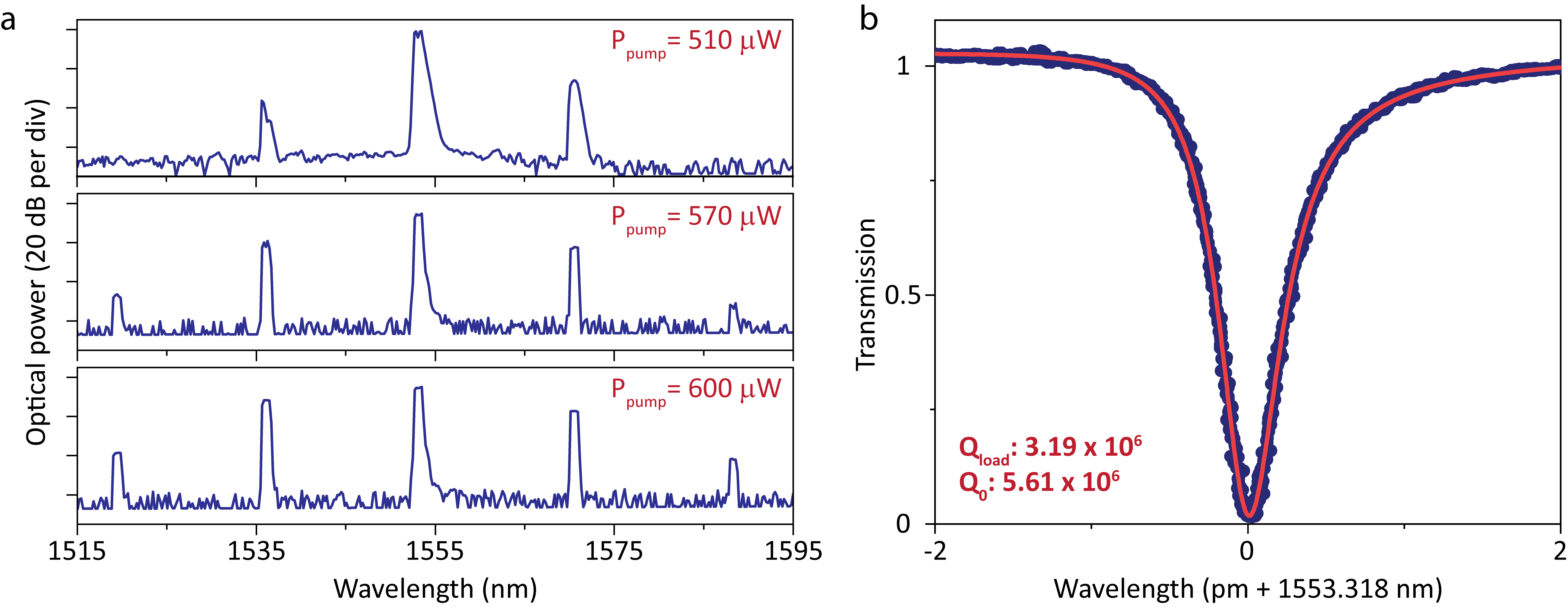}
\captionsetup{format=plain,justification=RaggedRight}
\caption{{\bf{Sub-mW parametric oscillation threshold power}} (\textbf{a}) SiC parametric oscillation induced by pumping at the wavelength of 1553.3~nm. Top panel shows OPO just above the threshold power (510~\textmu W total power in the waveguide). Middle and lower panels show measured optical spectra with loaded pump power of approximately 570 and 600~\textmu W, respectively. (\textbf{b}) High-resolution scan of the fundamental TE mode with a loaded (intrinsic) quality factor of 3.19~(5.61)~million. The mode is seen to be nearly critically-coupled to the waveguide. The scan laser wavelength is calibrated using a wavemeter, and the red curve is a fit to a Fano lineshape. The asymmetry of the resonance shape is attributed to interference with back-reflection of the vertical couplers. }
\label{fig:OPO_threshold}
\end{figure}

The efficiency of the Kerr nonlinear interaction improves with higher quality factors of the optical resonator. For example, the threshold relation for optical parametric oscillation in a microresonator can be expressed in the following form:
\begin{align}
P_{th} = \frac{\pi n \omega_{0}A_{\text{eff}}}{4\eta \ n_{2}}\frac{1}{D_{1}Q^2}
\end{align}
where Q denotes the total Q factor (intrinsic loss and loading included) with pump mode frequency $\omega_{0}$, $A_{\text{eff}}$ is the effective mode area, $\eta$ is the cavity-waveguide coupling strength, and $D_{1}$ is the free-spectral range (FSR) in units of rad/s. Zero detuning of the laser frequency with respect to the pump mode frequency is assumed. The parametric oscillation threshold is inversely proportional to the square of the Q factor. Fig.~\ref{fig:OPO_threshold} shows a sub-milliwatt (approximately 510 \textmu W) parametric oscillation threshold of a SiC optical resonator featuring an intrinsic Q factor of 5.6 million with a 350 GHz FSR.   

\subsection{Demonstration of a Silicon Carbide soliton microcomb}
\begin{figure}[h!]
\centering
\includegraphics[width=0.9\linewidth]{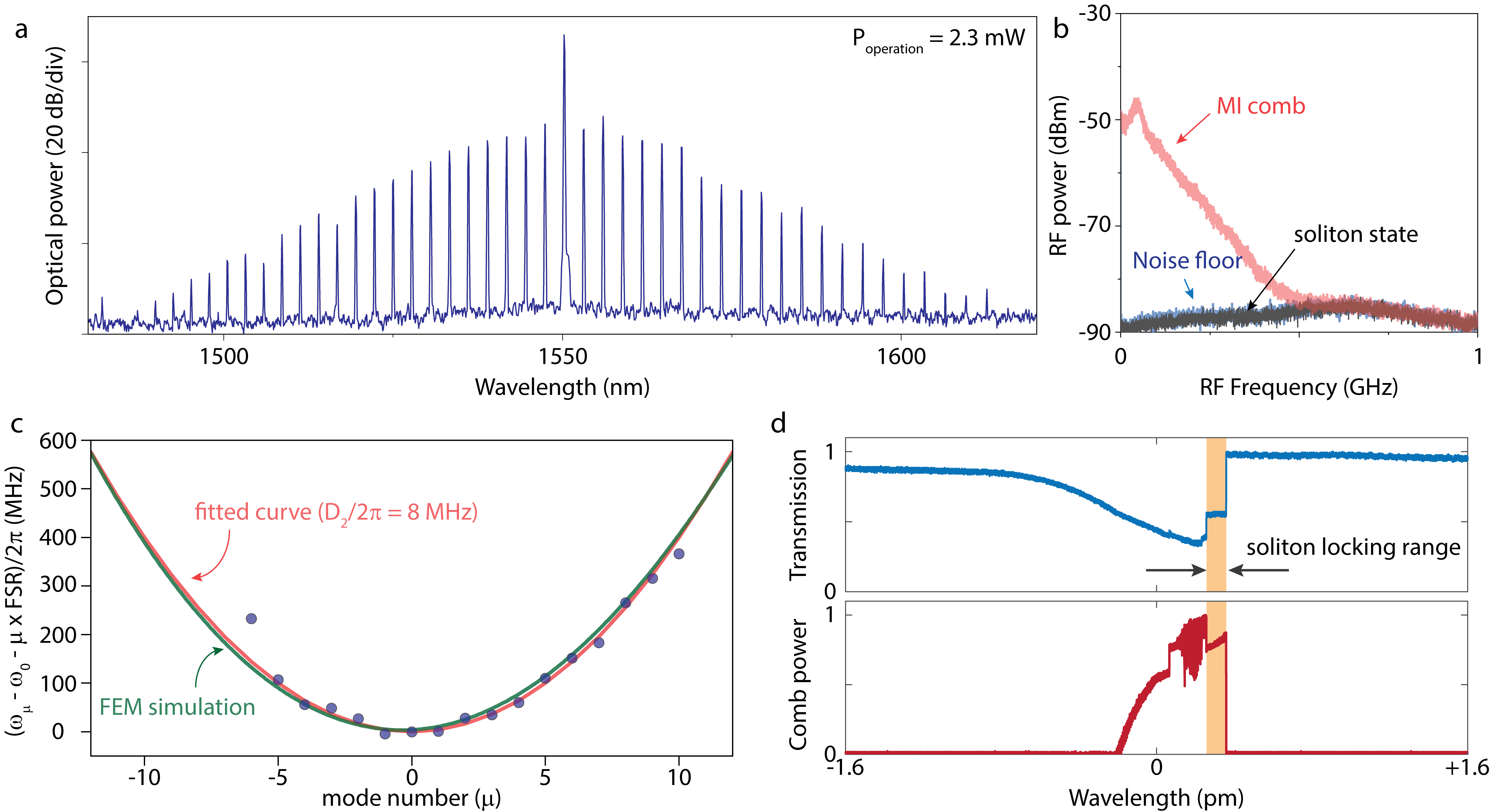}
\captionsetup{format=plain,justification=RaggedRight}
\caption{{\bf{SiC soliton microcomb}} (\textbf{a}) The optical spectrum of a single soliton state with 2.3~milliwatts operation power. (\textbf{b}) RF spectrum (resolution bandwidth = 100~kHz) of the entire soliton comb confirms a low-noise state. (\textbf{c}) Measured frequency dispersion belonging to the soliton forming mode family (TE$_{00}$) is plotted versus the relative mode number. The red curve is a fit using D$_{1}$/2$\pi$ = 358.663 GHz and D$_{2}$/2$\pi$ = 8 MHz. Simulation of the soliton mode families is plotted (green curve), and the simulation fairly agrees with the measurement results. (\textbf{d}) Upper panel presents pump power transmission versus tuning across a resonance used for the soliton formation. Lower panel shows comb power trace in which the pump laser scans over the resonance from the short wavelength (blue detuned) to the long wavelength (red detuned). The shaded region (orange) depicts the spectral region where the single soliton exists.}
\label{fig:single_soliton}
\end{figure}

Coherently pumped solitons in optical microresonators form as a result of the balance of the Kerr nonlinear shift and the cavity dispersion, as well as the parametric gain and the cavity loss. The soliton-forming mode family (in particular for bright solitons) in a microresonator must feature anomalous dispersion and minimal distortion of the dispersion (\textit{e.g.}, minimal avoided-mode-crossings). The power requirement for soliton operation is inversely proportional to the total Q factor of the mode family\cite{yi2015soliton}. 

\begin{table}[b!]\centering
\begin{tabular}{m{0.15\textwidth}>{\centering}m{0.15\textwidth}>{\centering}m{0.15\textwidth}>{\centering}m{0.35\textwidth}>{\centering\arraybackslash}m{0.15\textwidth}}
 \hline
 \textbf{Material} & \textbf{Q$_{0}$ (M)} & \textbf{FSR (GHz)} & \textbf{Soliton operation power \\(OPO threshold) (mW)} & \textbf{Reference} \\ 
 \hline\hline
 Si$_{3}$N$_{4}$ & 260 & 5 & $\sim$ 20 & \mbox{Ref [\!\!\citenum{Bowers:2021:NaturePhotonics}]} \\ 
 \\[-1em]
 Si$_{3}$N$_{4}$ & 8 & 194 & 1.3 (1.1) & \mbox{Ref [\!\!\citenum{Lipson:2018:Nature}]} \\ 
 \\[-1em]
 Si$_{3}$N$_{4}$ & 15 & 99 & 6.2 (1.7) & \mbox{Ref [\!\!\citenum{Kippenberg:2018:Optica}]} \\ 
 \\[-1em]
 SiO$_{2}$/Si$_{3}$N$_{4}$ & 120 & 15 & 28 (5) & \mbox{Ref [\!\!\citenum{Vahala:2018:NaturePhotonics}]}  \\ 
 \\[-1em]
 LiNbO$_{3}$ & 2.4 & 199.7 & 5.2 & \mbox{Ref [\!\!\citenum{Lin:2020:LPR}]} \\ 
 \\[-1em]
 AlGaAs & 1.5 & 450 & 1.77 (0.07) & \mbox{Ref [\!\!\citenum{Bowers:2020:NatureCommunications}]}  \\ 
 \\[-1em]
 SiC & 5.6 & 350 & 2.3 (0.51) & This work \\ 
 \\[-1em]
 \hline
\end{tabular}
\caption[Comparison of integrated soliton device performance]{\textbf{Comparison of integrated soliton device performance} }
\label{table:comparison}
\end{table}

We demonstrate the generation of a dissipative soliton microcomb in a SiC microresonator. Figure~\ref{fig:single_soliton}a shows the spectrum measured for a single-soliton state, and the soliton spectral shape follows the square of a hyperbolic secant function. Small spurs in the spectrum correlate with the avoided-mode-crossings in the mode dispersion spectrum (Fig.~\ref{fig:single_soliton}c), and the RF spectrum of the single-soliton state confirms that it is a low-noise state (Fig.~\ref{fig:single_soliton}b).  While tuning the laser through the resonance mode, the pump power transmission as well as the comb power (Fig.~\ref{fig:single_soliton}d) show a step transition from modulation instability (MI) and a chaotic comb state to a stable soliton comb state. The high Q SiC resonator enables a low operation power of the soliton microcomb of 2.3~mW: Table~\ref{table:comparison} compares operation powers of various chip-scale soliton devices. 

\section{Soliton crystals}
Soliton crystals, temporally-ordered ensembles of soliton pulses, have been observed in various optical resonator platforms, and their dynamics as well as defect-free generation have been actively explored. We demonstrate soliton crystal states with 2- and 7-FSR comb spacing, corresponding to phase-locked lattice of 2 and 7 identical solitons, respectively. We characterize the soliton crystals through the analysis of their optical spectra, RF beatnote, and second-order photon correlations. 

\subsection{2-FSR soliton crystal}
\begin{figure}[h!]
\centering
\includegraphics[width=0.9\linewidth]{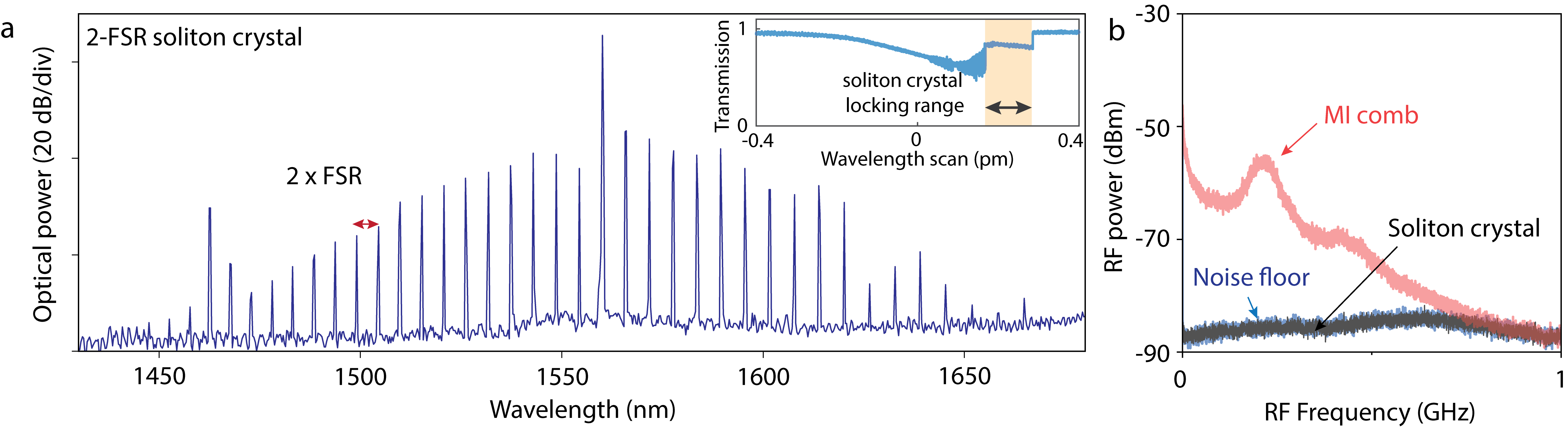}
\captionsetup{format=plain,justification=RaggedRight}
\caption{{\bf{2-FSR soliton crystal state}} (\textbf{a}) OSA spectrum of the soliton crystal state. Inset: pump power transmission versus laser tuning when the pump laser wavelength is scanned from blue- to red-detuned across the pump resonance. (\textbf{b}) RF spectra (resolution bandwidth = 100 kHz) of the soliton comb (black) and MI comb (red).  }
\label{fig:PSC2FSR}
\end{figure}

Figure~\ref{fig:PSC2FSR} shows the OSA spectrum and RF beatnote noise of the soliton crystal state that is studied in Fig.~5 of the main text. In the main text, the RF beatnote of a single resonator mode is shown; In Fig.~\ref{fig:PSC2FSR}, we show the RF spectrum of the whole comb. Sweeping the laser from blue- to red-detuned, we observed a transition from a broad and noisy RF signal corresponding to the modulation instability (MI) state, to a low-noise state, coinciding with the beginning of a discrete step in the transmission trace across the cavity resonance. %It was confirmed that the pump mode does not exhibit any resolvable mode splitting.

\subsection{Photonic molecule analysis}
The second-order correlation matrix for the 2-FSR soliton crystal state (presented in Fig.~5c of the main text) was computed via LLE simulation and input-output theory using the following parameters:
\begin{itemize}
    \item $D_2/2\pi = 3.65$~MHz, obtained from FEM simulation, neglecting higher-order terms. A single perturbation of $-30$~MHz was introduced at mode $\mu=-2$ to induce the formation of the soliton crystal state \cite{Kippenberg:2019:SolitonCrystals}.
    \item For the pump mode ($\mu = 0$), the intrinsic and coupling Q factors of 2.37 and 6.55 million, respectively, were used, extracted from the measured cold-cavity transmission spectrum.
    \item For the other modes, intrinsic and coupling Q of 2.77 and 7.47 million, respectively, were used, corresponding to the mean of the measured Q factors for the modes within the laser scanning range ($\mu=-3$ to $+14$). 
    \item Pump power of 6.6~mW in the waveguide, corresponding to the experimentally-measured value.
\end{itemize}

\begin{figure}[h!]
\centering
\includegraphics[width=0.9\linewidth]{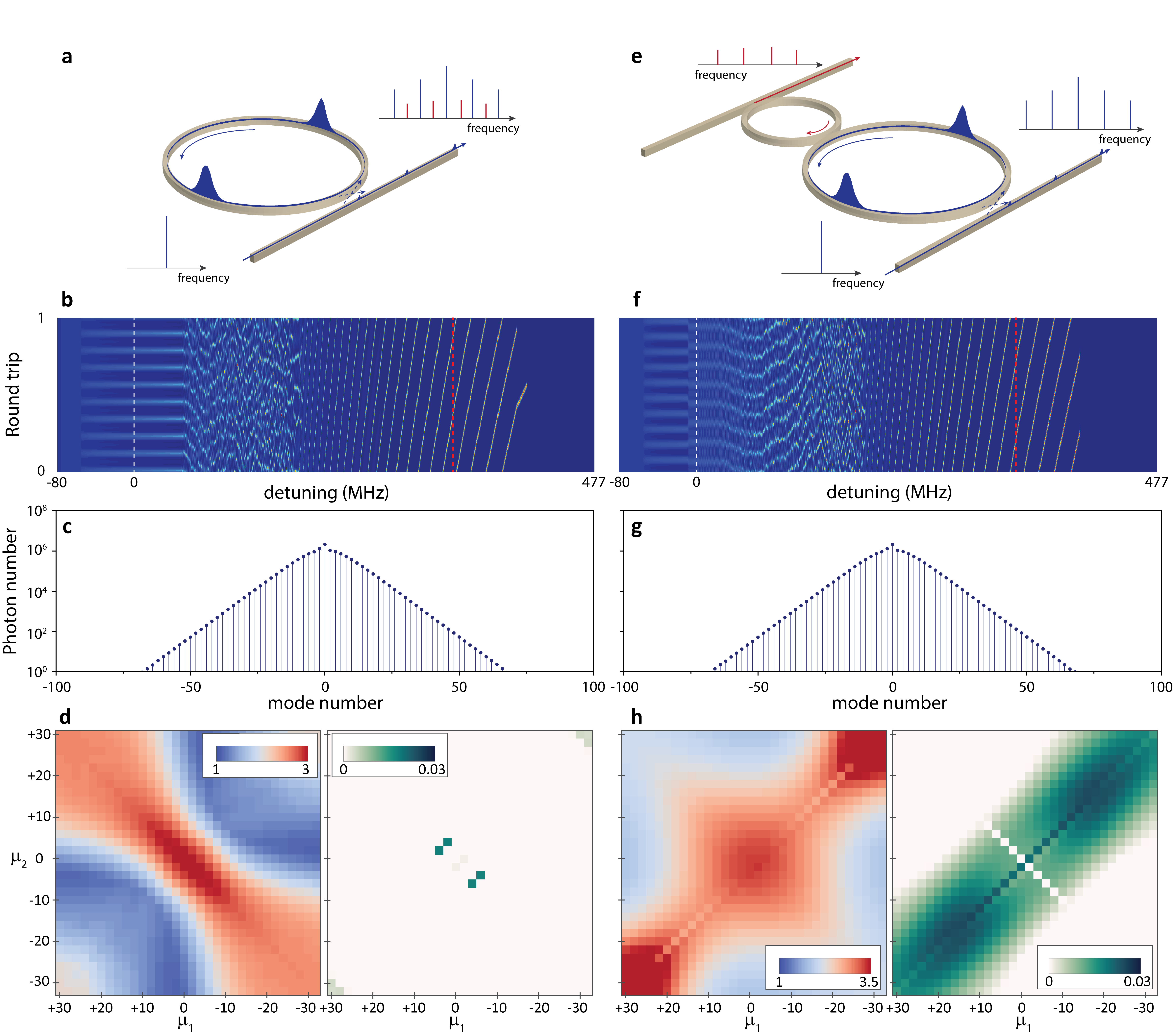}
\captionsetup{format=plain,justification=RaggedRight}
\caption{{\bf{Photonic molecule architecture for all-to-all entanglement generation}} \textbf{(a)} Schematic of the experimentally-demonstrated device. \textbf{(b)} LLE simulation of the device for pump power of 6.6~mW in the waveguide. The desired 2-FSR soliton crystal state exists for detuning in the range 175--395~MHz. This simulated soliton step width of 220~MHz is somewhat larger than the experimentally-observed step width of 150~MHz. \textbf{(c)} The simulated spectrum taken at detuning of 330~MHz. \textbf{(d)} Left: The two-photon correlation matrix computed for the state in (c). The scale bar indicates $\max\{g^{(2)}(\tau)\}.$ Right: The corresponding entanglement negativity, $E_N$, matrix. (e-h) correspond to (a-d) but for the photonic molecule configuration, where the out-coupling of the odd resonator modes is increased by 10 times via the auxilliary resonator \textbf{(e)} Schematic of the photonic molecule configuration. \textbf{(f)} It is confirmed via LLE that the same 2-FSR soliton crystal state can be captured in simulation. \textbf{(g)} The spectrum of the comb at the same detuning of 330~MHz is identical to the comb spectrum in the unmodified device, since only the below-threshold modes are affected by the addition of the auxiliary ring. \textbf{(h)} The corresponding correlation and entanglement matrices for the photonic molecule device.}
\label{fig:molecule}
\end{figure}

The result of the LLE simulation is shown in Fig.~\ref{fig:molecule}b. The simulated soliton crystal spectrum for the detuning of 330~MHz is shown in Fig.~\ref{fig:molecule}c. In the input-ouput theory model, the laser detuning (within the range of existence of the soliton state in the LLE simulation) is the only free parameter. The corresponding second-order photon correlations and $E_N$ matrices are shown in Fig.~\ref{fig:molecule}d. Negligible entanglement is thus predicted in the resonator mode basis for this soliton crystal state. However, entanglement can be recovered by selectively over-coupling the below-threshold modes via a photonic molecule configuration, shown in  Fig.~\ref{fig:molecule}e. This configuration is as follows: The auxillary resonator has a FSR that is 2 times larger than the FSR of the primary microring. The coupling strength of the two ring resonators exceeds the total losses (scattering and waveguide coupling) of the primary resonator. The auxillary ring is further over-coupled to its output waveguide, so that rather than be strongly-coupled to the primary resonator, the auxillary resonator acts as a selective out-coupling channel for the odd-numbered modes of the primary resonator. We note that the finesse of the experimentally demonstrated resonators (approximately 3500) is sufficient for this architecture. To model this system, we perform the LLE simulation with the same device parameters as for the experimentally demonstrated device, but with the out-coupling rates of the odd modes increased by 10 times. We numerically confirm that the same 2-FSR soliton state can be prepared for this device (Fig.~\ref{fig:molecule}f). The computed second-order correlation and $E_N$ matrices for the the quantum state of this device are shown in Fig.~\ref{fig:molecule}h. 

\clearpage

\subsection{7-FSR soliton crystal}
\begin{figure}[h!]
\centering
\includegraphics[width=0.9\linewidth]{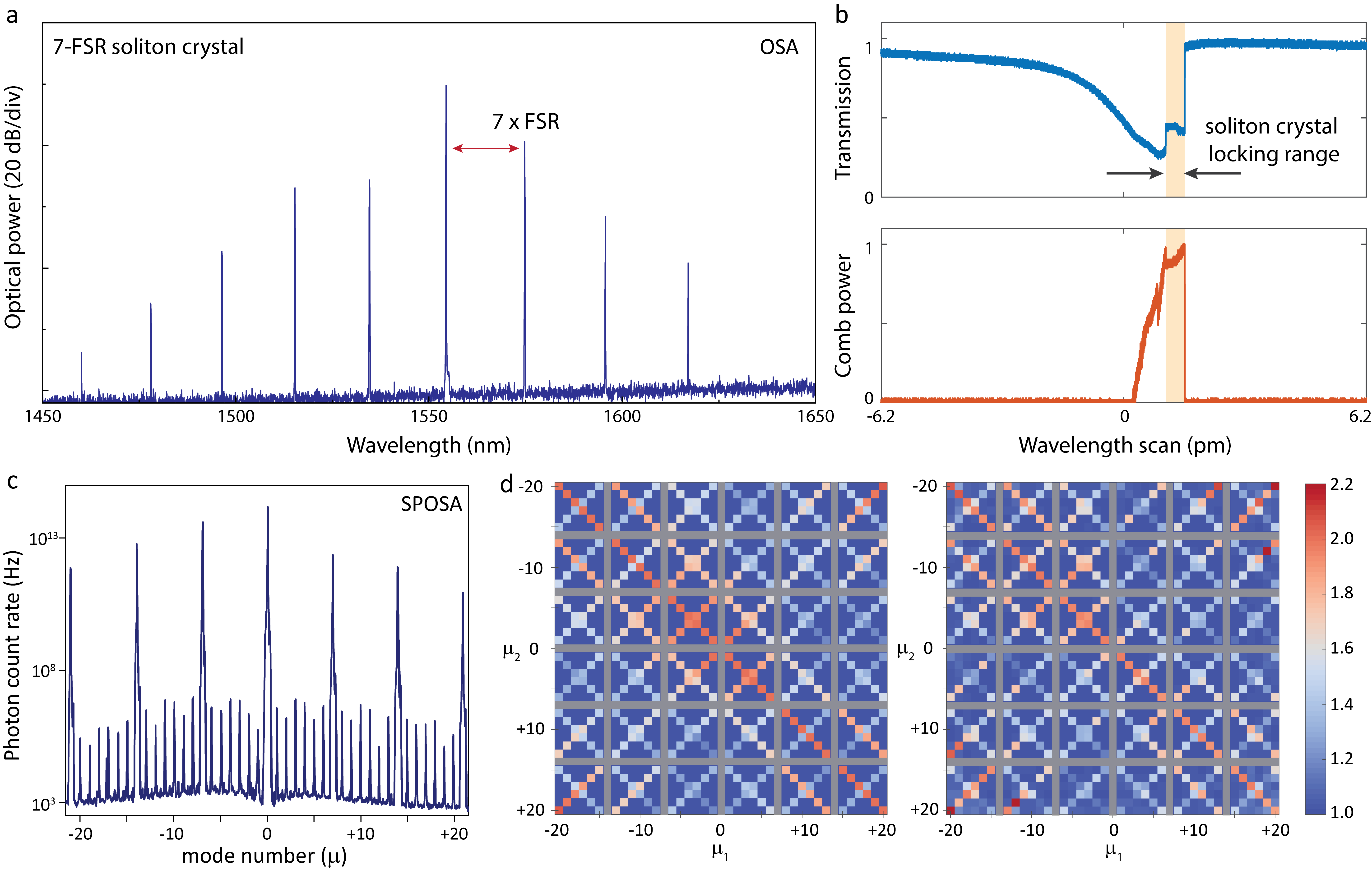}
\captionsetup{format=plain,justification=RaggedRight}
\caption{{\bf{7-FSR soliton crystal state}} (\textbf{a}) OSA spectrum of the soliton crystal state. (\textbf{b}) Pump power transmission (upper panel) and comb power (lower panel) versus wavelength tuning when the pump laser is scanned from blue to red across the pump resonance. (\textbf{c}) Optical spectrum of the soliton state measured using the SPOSA. (\textbf{d}) The $\text{max}[g^{(2)}(\tau)]$ correlation matrix for the below threshold modes in the 7-FSR soliton state (Left: theoretical model, Right: experimental data). }
\label{fig:PSC7FSR}
\end{figure}

Figure~\ref{fig:PSC7FSR} presents the generation of a 7-FSR soliton crystal state in a different device.  The optical spectrum (Fig.~\ref{fig:PSC7FSR}a) as well as the transmission and comb power traces across the pump resonance identify the existence of the soliton state. 
The SPOSA spectrum (Fig.~\ref{fig:PSC7FSR}c) reveals quantum frequency comb lines which were obscured by the noise floor of the OSA spectrum, and their correlation matrix is presented with the prediction from the LLE-driven linearized model (Fig.~\ref{fig:PSC7FSR}d).        

\bibliography{supplement}